\documentstyle[sprocl]{article}

\def\Journal#1#2#3#4{{#1} {\bf #2}, #3 (#4)}

\def\NPB{{\em Nucl. Phys.} B}
\def\PLB{{\em Phys. Lett.}  B}
\def\EPJC{{\em Eur. Phys. Journal} C}

\def\PRD{{\em Phys. Rev.} D}

\def\be{\begin{equation}}
\def\ee{\end{equation}}
\def\bea{\begin{eqnarray}}
\def\eea{\end{eqnarray}}

\begin{document}

\rightline {LPHEA/00-02}
\rightline {UFR-HEP/00-12}
\rightline {August 2000}

\title{Comments On the Confinement from Dilaton-Gluon Coupling in QCD}
\author{M. Chabab}
\address{LPHEA, Physics Department, Faculty of Science, Cadi Ayyad University,\\
P.O. Box 2390, Marrakesh 40001,
Morocco\\E-mail:mchabab@ucam.ac.ma}
\address{ and UFR-HEP, Department of Physics, Faculty of Science,Av. Ibn Battouta,\\
P.O. Box 1014, Rabat, Morocco}
\maketitle\abstract{In this talk, I report on a work done in
collaboration with R. Markazi and E.H. Saidi [1].}
\section{Introduction}
{Confinement in gauge theories provides one of the most
challenging problems in theoretical physics. Various quark
confinement models rely on flux tube picture. The latter emerges
through the condensation of magnetic monopoles and explain the
linear rising potential between color sources. However, a deep
understanding of confinement mechanism in still lacking.\\
Recently it has been shown in [2] that a string inspired coupling
of a dilaton $\phi$ to the 4d $ SU(N_c)$ gauge fields yields a
phenomenologically interesting interquark potential $V(r)$ with a
confining term. Extension of gauge field theories by inclusion of
dilatonic degrees of freedom has gained considerable interest.
Particularly, Dilatonic Maxwell and Yang Mills theories which,
under some assumptions, possess stable, finite energy solutions
[3].}
\section{Description of Dick Model}
The Dick interquark potential [2] was obtained as follow: First
start from the following model for the dilaton gluon coupling
$G(\phi)$:
\begin{equation}
 L(\phi,A)=-{1\over 4G(\phi)} F_{\mu \nu}^{a}F_{a}^{\mu \nu}-{1\over2}(\partial_{\mu}{\phi})^2+W(\phi)+J_{\mu}^{a}A_{a}^{\mu}\qquad
\label{eq:quantum}
\end{equation}
Then construct $G(\phi)$ under the requirement that the Coulomb
problem still possesses analytical solutions.The coupling
$G(\phi)$ and the potential $W(\phi)$ that emerged are:
\begin{equation}
G(\phi)=const.+{f^2\over \phi^2},\qquad 
W(\phi)={1\over 2}m^2\phi^2 \qquad
\label{eq:quantum}
\end{equation}
where $f$ is a scale parameter characterizing the strength of the
scalar-gluon coupling and $m$ represents the dilaton mass.

Next, consider the equations of motion of the  fields $A_\mu$ and
$\phi$  and solve them for static point like color source
described by the current density $J_a^{\mu}=\rho_a\eta^{\mu 0}$.
After some straightforward algebra, Dick shows that the
interquark potential $V_{D}(r)$ is given by (up to a color factor),
\begin{equation}
V_D(r)={\alpha_s \over r}-gf\sqrt{N_c\over{2(N_c-1)}}\ln[exp(2mr)-1]\qquad
\label{eq:quantum}
\end{equation}
Eq.(3) is remarkable since for large values of $r$ it leads to a
linear confining potential
$V_D(r)\sim2gfm\sqrt{N_c\over{2(N_c-1)}}r$.

This derivation provides a challenge to monopole condensations as a
new  quark confinement scenario. Therefore, it is justified to
dedicate more efforts to the investigation of a more general
effective coupling function $G(\phi)$ and to the phenomenological
application of Dick potential $V_{D}(r)$ [5].

In this regard, we have shown in [1] that for a general dilaton-gluon coupling $G(\phi)$,
 the quark interaction potential $V(r)$ reads as:
\begin{equation}
V(r)= \int dr{G[\phi (r)]\over {r^2}}\qquad  \label{eq:quantum}
\end{equation}
Such form of the potential is very attractive. On the one hand, it
extends the usual Coulomb formula $V_c \sim 1/r$ which is
recovered from (4) by taking  $G=1$. Moreover for $G \sim r^2$,
which by the way corresponds to a coupling
$G(\phi)\sim~\phi^{-2}$, and $W(\phi)={m^2\over 2}\phi^2$,  $m \ne
0$, Eq.(4) yields Dick solution.
On the other hand Eq.(4) may be also used to relate non
perturbative effects such as QCD vacuum condensates in term of
dilaton parameter $(m, f)$. Indeed, following for instance[4],
one may extract interesting phenomenological informations on the
dilaton-gluon coupling $G[\phi]$ by comparing Eq.(4) to the
Bian-Huang-Shen's potential $V_{BHS}(r)$ namely:
\begin{equation}
V_{BHS}(r)\sim {1\over r}-{\sum_{n\ge 0}C_nr^n}\qquad
\label{eq:quantum}
\end{equation}
where $C_n$'s are related to the quark and gluon vacuum
condensates. In fact one can do better if one can put the coupling
$G(\phi)$ in the form $G[\phi(r)]$. In this case one can predict
the type of vacuum condensates  of the $SU(N_c)$ gauge theory
which contributes to the quark interaction potential.\\

Although, the derivation of the formula (4) for the interquark
potential from Eq.(1) is by itself an important result, there
remain however other steps to overcome before one can exploit (4).
A crucial technical step is to determine for what type of
couplings $G(\phi)$, one can solve the
equation of motion of the scalar field $\phi$:
\begin{equation}
[D_{\mu},G^{-1}(\phi)F^{\mu\nu}]=J^\nu \quad(a)\quad,\qquad 
\partial_\mu \partial^\mu \phi={\partial W\over \partial \phi}- {1\over 4}F_{\mu \nu}^a F_a^{\mu\nu}{\partial G^{-1}(\phi)\over \partial
\phi} \quad(b)
\label{eq:quantum}
\end{equation}
In trying to explore (4), we hav observed that the functional $G[\phi(r)]$, and then the potential V(r) of (4) may be
obtained from the following one dimensional lagrangian 
\begin{equation}
L_D={1\over 2}(y')^2+r^2W(y/r)+{\alpha\over {2r^2}}G(y/r)\qquad
\label{eq:quantum}
\end{equation}
where $y=r\phi$, $y'=({dy\over dr})$and $\alpha={g^2\over
{16\pi^2}}{{N_c-1}\over{2N_c}}$ and where $g$ is the gluon
coupling constant. Indeed, if we start from (6) and set 
$F_a^{0i}=-{gC_a\over4\pi}\partial_i V$, the equations of motion
take a simple form,
\begin{equation}
{dV\over dr}=r^{-2}G[\phi]\qquad  (a)\quad,\quad
\Delta\phi={\partial W\over \partial\phi}+{\alpha\over
r^4}{\partial G(\phi)\over\partial\phi}\qquad (b)
\label{eq:quantum}
\end{equation}
Eq.(8.b) can be interpreted as corresponding to a mechanical system
with the action:
\begin{equation}
S=\int dr{ r^2[{\phi}^2+W(\phi)+{\alpha\over {2r^4}}G(\phi)]}\\
    = \int dr{[{1\over 2}(y')^2+r^2W(y/r)+{\alpha\over {2r^2}}G(y/r)]}
 \label{eq:quantum}
\end{equation}
Consequently the coupling $G(\phi)$ of
Eq.(1) appears as a part of interacting  potential of $1d$
quantum field theory. 
\section{Genaralized Dick model}
\qquad First of all observe that the lagrangian (7) including the
Dick model (1) is a particular one dimensional field theory of
lagrangian
\begin{equation}
L={1\over 2}(y')^2-U(y,r) \label{eq:quantum}
\end{equation}
where $U(y,r)$ is a priori an arbitrary potential. Though simple,
this theory is not easy to solve except in some special cases. A
class of solvable models is given by potentials of the form :
\begin{equation}
U(y)= \lambda^2 y^{2(n+p)}+\gamma^2 y^{2(q-n)}+\delta y^k
\label{eq:quantum}
\end{equation}
where $n, p, q$ and $k$ are numbers and $\lambda^2$ , $\gamma^2$
and $\delta$ are coupling constants scaling as $(lenght)^{-2}$.
The next thing to note is that Eq.(11) has no explicit dependence
in $r$ and consequently the following  identity usually holds :
\begin{equation}
y'^2 = U+c \label{eq:quantum}
\end{equation}
where c is  a constant. Actually Eq.(12) is just an integral of
motion which may be solved  under some assumptions. Indeed by
making appropriate choices of  the coupling $\lambda$ as well as
the integral constant c, one may linearise y' in Eq.(12) as
follows :
\begin{equation}
y' = U_1+U_2\qquad. \label{eq:quantum}
\end{equation}
Once the linearisation in y' is achieved and the terms $U_1$ and
$U_2$ are identified, we can show that the solutions of Eq.(12)
are classified by the product $U_1U_2$ and the ratio $U_1/U_2$. In
what follows we discuss briefly some interesting examples.

\subsection{First case: Dick solution}
\noindent This corresponds to take $U_1=my$ and
$U_2=c_1y^{-1}$. Putting back into Eq.(12) one gets the Dick
solution [2] which yields to the potential of Eq.(3).

\subsection{Second case: New solutions}
\noindent In this case the mass term  is related to the product $U_1U_2$  as:
\begin{equation}
U_1U_2=\pm {1\over 2}m^2y^2
\label{eq:quantum}
\end{equation}
Eq.(14) cannot however determine $U_1$ and $U_2$ independently
as in general the following realizations are all of them
candidates,
\begin{equation}
U_1=\lambda y^{n+p}, \qquad\qquad  U_2=\gamma y^{q-n}
\label{eq:quantum}
\end{equation}
where the integers $p$ and $q$ are such that $p+q=2$ and where
$\lambda \gamma=\pm m^2$. A remarkable example corresponds to take
$p=q=1$. In this case we distinguish two solutions according to
the sign of the product of $\lambda \gamma$. For $\lambda
\gamma=+m^2$, the solution is
\begin{equation}
y(r)=[{1\over \lambda} tan({nmr\over \sqrt2}+const.)]^{1\over
n}\qquad . \label{eq:quantum}
\end{equation}
For $\lambda\gamma=-m^2$, we have:
\begin{equation}
y(r)=[-{1\over \lambda} tanh({nmr\over \sqrt2}+const.)]^{1\over
n}\qquad. \label{eq:quantum}
\end{equation}
Note that one can go
beyond the above mentioned solutions which are just special cases
of general models involving interactions classified according to the
following constraint:
\begin{equation}
U_1.U_2\sim y^k \label{eq:quantum}
\end{equation}
with $k(=p+q)$ an integer. Indeed, besides $k=0$ and $k=2$ which
lead respectively to Dick solution and to the solutions given by 
(16,17); for general values of  $k $, one has
to know moreover the ratio $U_1/U_2$ in order to work out
solutions. For the example where
$$ U_1=\lambda y$$
\begin{equation}
U_2=\gamma y^{k-1}\qquad;\quad k\quad integer \label{eq:quantum}
\end{equation}
one can check, after some straightforward algebra, that the
solution 
\begin{equation}
y_k(r)=[r\phi_D]^{2\over (2-k)}\qquad . \label{eq:quantum}
\end{equation}
is just a generalization of Dick solution (case $k=0$).

\section*{Acknowledgments}
{The author wishes to thank Prof. Wolfang Lucha and Dr. Nora
Brambilla for their kind
invitation to the International Conference "ConfineIV" and for their hospitality.
This work has been partially supported by the Rectorat of Cadi Ayyad University.}

\end{document}